\documentclass{article}
\usepackage{spconf}

\usepackage{amsmath, bbm}
\usepackage{amssymb}
\usepackage{amsfonts}
\usepackage{amsthm}
\usepackage{algorithm}
\usepackage{algorithmic}
\usepackage{nicefrac}
\usepackage{bm}
\usepackage{hyperref}
\usepackage{cite}
\usepackage[utf8]{inputenc} 
\usepackage[T1]{fontenc}
\usepackage{graphicx}
\usepackage[font=small]{caption}
\usepackage{tikz}
\usetikzlibrary{positioning}
\usepackage{pgfplots}
\usepackage{epstopdf}
\usepackage{subcaption}
\usepackage{multirow}
\usepackage{lipsum}

\usepackage{tcolorbox}

\newtheorem{Problem}{Problem}

\input{mysymbol.sty}

\title{Adaptive Contention Window Design using Deep Q-learning}

\name{Abhishek Kumar$^\star$, Gunjan Verma$^\dag$, Chirag Rao$^\dag$, Ananthram Swami$^\dag$, and Santiago Segarra$^\star$
\thanks{Research was sponsored by the Army Research Office and was accomplished under Cooperative Agreement Number W911NF-19-2-0269. 
		The views and conclusions contained in this document are those of the authors and should not be interpreted as representing the official policies, either expressed or implied, of the Army Research Office or the U.S. Government. 
		The U.S. Government is authorized to reproduce and distribute reprints for Government purposes notwithstanding any copyright notation herein.
		\newline
		Emails:  \{ak109, segarra\}@rice.edu, \{gunjan.verma.civ, chirag.r.rao.civ, ananthram.swami.civ\}@mail.mil. }}
\address{$^\star$Rice University, USA  \hspace{1cm} $^\dag$US Army’s CCDC Army Research Laboratory, USA}

\begin{document}
\ninept
\renewcommand{\baselinestretch}{0.95}
\maketitle
\begin{abstract}
We study the problem of adaptive contention window (CW) design for random-access wireless networks.
More precisely, our goal is to design an intelligent node that can dynamically adapt its minimum CW (MCW) parameter to maximize a network-level utility knowing neither the MCWs of other nodes nor how these change over time.
To achieve this goal, we adopt a reinforcement learning (RL) framework where we circumvent the lack of system knowledge with local channel observations and we reward actions that lead to high utilities.
To efficiently learn these preferred actions, we follow a deep Q-learning approach, where the Q-value function is parametrized using a multi-layer perceptron.
In particular, we implement a rainbow agent, which incorporates several empirical improvements over the basic deep Q-network.
Numerical experiments based on the NS3 simulator reveal that the proposed RL agent performs close to optimal and markedly improves upon existing learning and non-learning based alternatives.
\end{abstract}
\section{Introduction}\label{sec:intro}
Wireless networks have become an essential part of our lives. 
We use them for a wide range of purposes including video streaming, web services, and file sharing.
Moreover, the number of wireless devices is constantly increasing, thus precipitating the urgent need for smart and efficient access to the limited available spectrum. 
Though oftentimes preferred due to their simplicity, conventional random access protocols such as IEEE 802.11 lead to suboptimal spectrum utilization and, more importantly, the performance of these protocols drops significantly under dynamic and uncertain settings~\cite{bianchi2000,bensaou2000fair,abyaneh2019intelligent}.

In this context, there has been growing interest over the last few years in improving the management of wireless networks under uncertain scenarios.
A testament to this trend is the recently completed DARPA spectrum collaboration challenge~\cite{SC2}.
In this competition, multiple wireless networks are present in the same geographical area and, without full information about the protocols employed by each network, they must interact with each other to efficiently utilize the shared spectrum. 
Furthermore, recent works on heterogeneous networks~\cite{yu2019heterogeneous,yu2019heterogeneous2,tan2019deep} analyze the efficient spectrum management in the case where nodes in the \emph{same} network may follow different access protocols.

Following this trend, we consider the scenario where nodes \emph{do} know the protocol followed by others but the uncertainty stems from not knowing key parameters governing this protocol.
In particular, we consider a binary exponential backoff protocol with the minimum contention window (MCW) being the key unknown parameter to be designed.
The proposed uncertain scenario is of empirical relevance in cases where nodes may act unfairly by reducing their MCW to acquire more channel access at the expense of others.
In this setting, we would rely on intelligent agents to adapt their own MCW to restore fairness~\cite{abyaneh2019intelligent}.
A similar situation might arise when a new node joins a network and, without full knowledge of others' parameters, the new node seeks to learn their dynamics and predict the optimal MCW value at every time. 
We address this setting of \emph{parameter uncertainty} by modeling the problem through a reinforcement learning (RL) framework and adapting state-of-the-art deep learning techniques for its solution.

\vspace{1mm}\noindent {\bf Related work.}
The problem of optimal CW design has been studied extensively over the past decades~\cite{syed2016adaptive,xia2006contention,chen2010random, chun2012adaptive,hong2012channel,deng2008contention,ksentini2005determinist}. 
For example, authors in~\cite{syed2016adaptive} and~\cite{chun2012adaptive} set the optimal MCW as a linear function of the estimated number of active nodes. 
Furthermore, control-based~\cite{xia2006contention} and game-theoretic~\cite{chen2010random} approaches have been developed to avoid the need for estimating the number of nodes.
However, unlike our setting, all these approaches assume that the nodes behave in a cooperative manner by either choosing the same MCW or by not deviating from a prespecified behavior.
Another class of approaches relies on modifications in the increase and decrease of the CW instead of finding an optimal MCW~\cite{bensaou2000fair, bharghavan1994macaw, song2003eied}.
For instance, under~\cite{bharghavan1994macaw}, a node multiplies its current CW by a constant factor if packets collide and decreases it by subtracting a constant value if the transmission is successful.
Our approach differs from this body of work since it does not deviate from the current protocol structure and is learning-based as opposed to being controlled by preset rules.

Motivated by the recent success of machine learning applied to wireless networks \cite{naparstek2017deep,wang2018deep,luong2019applications,ye2019deep,nasir2019multi,chowdhury2020unfolding}, learning-based solutions have been proposed for optimal distributed CW design~\cite{edalat2019dynamically, pressas2019qlearningcontention, abyaneh2019intelligent,ali2017machine,amuru2015send}. 
Germane to our approach is~\cite{pressas2019qlearningcontention}, where (non-deep) Q-learning is applied. 
However, since the Q-function is learned as a table instead of parametrized by a neural network, the authors deviate from the current protocol and consider a different (much smaller) state-action space to enable learning.
Another relevant approach is presented in~\cite{abyaneh2019intelligent}, which considers a similar setting to ours but proposes a supervised learning strategy based on a random forest algorithm. 
As such, it performs well in a static scenario but fails to generalize to the cases where other nodes vary their CW parameters, as we illustrate through experiments.

\vspace{1mm}\noindent\textbf{Contribution.} 
The contributions of this paper are twofold:\\
i) We formulate the problem of adaptive MCW design as an RL problem and implement a deep Q-learning architecture for its solution.\\
ii) Through NS3 simulations, we compare the proposed solution with established baselines and demonstrate its near-optimal performance across several scenarios.

\section{System Model and Problem Statement}
\label{sec:problem}


We consider a wireless network with $N$ nodes transmitting packets to a central access point. 
Nodes follow a binary exponential backoff mechanism as mentioned in the IEEE 802.11 protocol for channel access. The exponential backoff mechanism reduces collisions in the channel by specifying the amount of time slots that each node must wait before a transmission~\cite{bianchi2000}. 
More precisely, before every transmission, a node draws the number of slots to wait from a uniform distribution in $(0,c-1)$, where $c$ is equal to the MCW $\omega$ if the previous transmission was successful.
After the first collision, $c$ is increased to $2 \omega$ to promote longer waiting times and reduce the probability of further collisions. 
In general, $c$ is chosen as $2^{j} \omega$ after $j$ successive collisions, and saturates at a maximum permissible prespecified value. 
The network is assumed to be operating in a high-load regime where every node has packets to transmit at all times. 
Furthermore, all nodes are assumed to be in range of all other nodes and there are no hidden nodes~\cite{bianchi2000}.

The network is observed in discrete intervals $k = 1, 2, \ldots$, each of duration $T$ seconds. 
Within each time interval $k$, we assume the MCW of all nodes $\bbomega^k = [\omega^k_0, \omega^k_1, \ldots, \omega^k_{N-1}]^\top$ to be constant. 
As described in Section~\ref{sec:intro}, we consider the problem of CW design from the point of view of an intelligent node (here node $0$) that seeks to optimize $\omega_0^{k}$ without knowledge of the MCW of other nodes and, even more challenging, in a setting where these values might change from one interval $k$ to the next.
We model the MCW of every other node $\omega_i^k$ for $i \neq 0$ as a discrete-time stochastic process that can take values in a finite state space $\Omega$.
In selecting the right MCW, node $0$ must rely exclusively on local information.
In particular, for each interval $k$, apart from knowing its own $\omega_0^k$, node $0$ observes the fraction of time $f^k$ that it transmits without collision and the fraction of time $b^k$ that other nodes transmit without collision.
Based on this, we define the observation vector $\bbo^k = [f^k, b^k]^\top$. 
Notice that $\bbo^k$ is a random vector that depends on $\bbomega^k$ through the (stochastic) exponential backoff mechanism.
Lastly, to discuss \emph{optimal} CW design we need to focus on a notion of utility.
In our case, we associate interval $k$ with the fairness utility given by
\begin{equation}\label{E:utility}
    u(\bbo^k, N) = 1 - \left|\frac{f^k}{b^k+f^k} - \frac{1}{N}\right|.
\end{equation}
Intuitively, with $N$ nodes in the network and under the high-load assumption, the fair share of node $0$ would be to transmit $1/N$-th of the total transmit time during interval $k$, while the actual proportion of transmit time is given by ${f^k}/({b^k+f^k})$.
The discrepancy between these two numbers can be seen as a fairness loss and the utility $u(\bbo^k, N)$ subtracts this loss from a perfect utility of $1$.
Notice that in computing~\eqref{E:utility}, the total number of nodes $N$ is assumed to be known to (or accurately estimated by) node $0$.
With this notation in place, we formally state our optimal CW design problem.

\begin{Problem}\label{P:main}
	Given $N$ and the history $\{\bbo^k, \omega_0^k\}_{k=1}^{K-1}$, determine the MCW $\omega^{K}_0$ to maximize $\mathbb{E} \left( \sum_{k=K}^\infty \alpha^{k-K} u(\bbo^k, N) \right)$.	
\end{Problem}

At every discrete-time instant $K$, our intelligent node is faced with an instance of Problem~\ref{P:main}: it knows the size of the network $N$, its past decisions $\{\omega_0^k\}_{k=1}^{K-1}$, and how these past decisions worked out in terms of realized proportions of channel occupancy $\{\bbo^k\}_{k=1}^{K-1}$. 
With this information, node $0$ seeks to maximize future fairness. 
In general, it seeks to maximize not only the immediate fairness at time $K$ but rather an infinite-horizon discounted sum, where $\alpha \in (0,1)$ establishes the relative importance of future utilities.
Since $\bbo^k$ is a random vector, future utilities are random variables, as is their discounted sum.
Hence, the objective of node $0$ is to maximize the expected value of this discounted sum.
In the next section, we frame Problem~\ref{P:main} from an RL perspective and propose to solve it via deep Q-learning.
Finally, notice that even though in this paper we focus on fairness, a similar problem can be postulated to maximize throughput or other measures of interest by modifying the utility in~\eqref{E:utility}.

\section{Deep Q-Learning for CW design}
\label{sec:solution}

A first naive approach towards solving Problem~\ref{P:main} would be to try to explicitly compute the expected value of the discounted utilities.
The first obstacle in achieving this is that we do not have access to the underlying stochastic process regulating the evolution of $\omega_i^k$ for $i \neq 0$.
Thus, we would first need to adopt a model for this stochastic process -- e.g., a Markov process with $|\Omega|^{N-1}$ states representing the $|\Omega|$ possible choices of the $N-1$ nodes -- and then estimate from data the parameters of this model such as the transition probability matrix of the Markov process.
Apart from being computationally expensive, we would then face a second obstacle related to partial observability, namely that node $0$ does not have access to the true state of the network $\bbomega^k$ but rather to a real-valued observation $\bbo^k$ whose distribution depends on $\bbomega^k$. 
Inferring the transitions between these hidden states further increases the complexity, indicating that a classical model-based approach is not well-suited to solve Problem~\ref{P:main}.

We turn our attention to a learning-based approach under the framework of RL, where the lack of system knowledge is overcome through methodical interactions with the system.
In particular, inspired by its success on other problems pertaining to wireless networks~\cite{naparstek2017deep,wang2018deep,luong2019applications,ye2019deep,nasir2019multi}, we focus on Q-learning. 
RL problems are modelled using Markov decision processes where an agent learns by interacting with an environment. 
At every time step $k$, the agent takes an action $a_{k}$ that affects the state of the environment. 
The state changes from $s_{k}$ to $s_{k+1}$ and the agent gets a reward $r(s_{k},a_{k})$.
The action taken by the agent at each state $a_k = \pi (s_k)$ is given by the policy $\pi$, a mapping from the set of states $\ccalS$ to the set of actions $\ccalA$. 
The goal of Q-learning is to find an optimal policy $\pi^*$ that maximizes the long-term expected accumulated discounted reward.
To do this, let us define the optimal Q-function $Q^*: \ccalS \times \ccalA \to \reals$ where the value $Q^*(s,a)$ corresponds to the long-term reward of selecting action $a$ in state $s$ and, from that point onward, following the optimal strategy $\pi^*$.
From this definition, it follows that 
\begin{equation}\label{E:opt_strategy}
\pi^*(s) = \argmax_{a} Q^*(s,a).
\end{equation}
In our setting, the RL agent is the intelligent node $0$, the action space is given by the MCW choices, the reward is given by the utility in~\eqref{E:utility}, and we model the state as an $M$-memory buffer (or history) of the local observations at node $0$, i.e. $\{\bbo^k, \omega_0^k\}_{k=K-M+1}^K$.
Having drawn this analogy, it follows that if we can compute $Q^*$ then the solution of Problem~\ref{P:main} is given by the optimal strategy $\pi^*$ as in~\eqref{E:opt_strategy}.

The classical way of finding $Q^*$ is through a value iteration method based on the Bellman equation~\cite{sutton1998introduction}
\begin{equation}\label{E:value_iteration}
    Q(s,a) \leftarrow Q(s,a) + \eta \left( r(s,a) + \alpha \max_{a'} Q(s',a') - Q(s,a) \right),
\end{equation}
where $\eta$ is a pre-defined learning step-size, $\alpha$ is the discount factor in Problem~\ref{P:main}, and $s'$ is the realized state after $s$.
It has been shown~\cite{sutton1998introduction} that under certain conditions on the step $\eta$ and the frequency of visits to the different states, the generic Q-function in~\eqref{E:value_iteration} is guaranteed to converge to $Q^*$, from where $\pi^*$ can be obtained [cf.~\eqref{E:opt_strategy}], seemingly solving Problem~\ref{P:main}.
However, the convergence to $Q^*$ can be extremely slow in practice.

Classical Q-learning performs well in settings where the state-action space is small, since the table-like iterative procedure in~\eqref{E:value_iteration} updates each of the $|\ccalS| \times |\ccalA|$ pairs separately.
By contrast, in our setting we have an agent (node $0$) whose action space can potentially be large (different choices for the MCW) and, more critically, the observations $\bbo^k$ from the environment are real-valued.
To extend the benefits of Q-learning to our more challenging scenario we rely on a \emph{deep} Q-learning framework~\cite{mnih2015human}, where we parameterize the Q-function as a neural network and update its parameters as opposed to independent entry-wise updates of the Q-function.
More precisely, with $\theta$ representing the parameters of our neural network -- usually called a deep Q-network (DQN) -- we seek to determine the optimal parameters $\theta^*$ such that $Q_{\theta^*}(s,a) \approx Q^*(s,a)$.

\begin{figure}[t]
	\centering
    \includegraphics[width=0.77\linewidth]{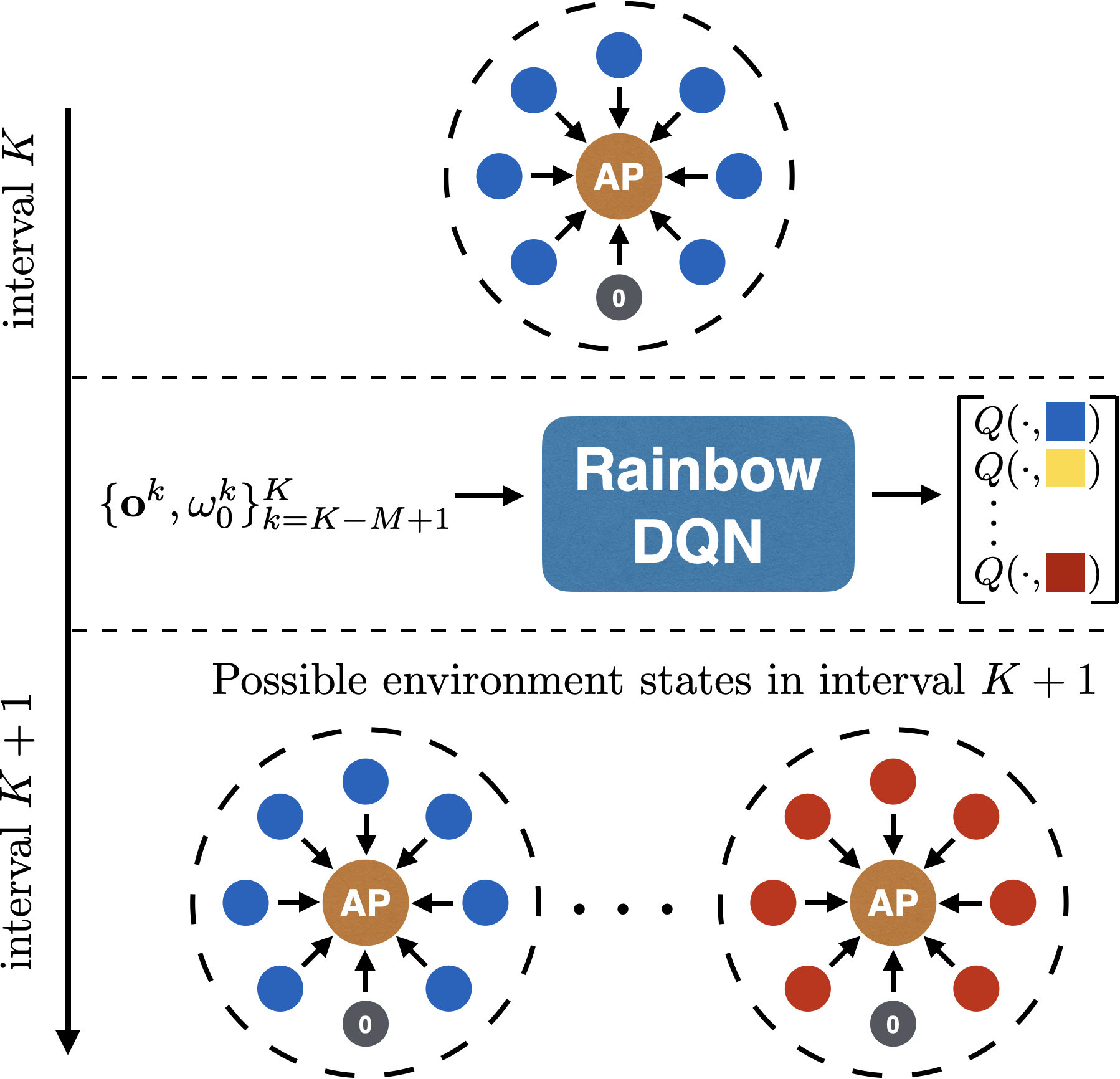}
	\caption{Schematic view of our proposed deep RL agent. Node $0$ relies on a rainbow DQN to select the MCW for the next time interval based on historical local observations.}
	\label{fig:main}
\end{figure}

The specific implementation of deep Q-learning for the solution of Problem~\ref{P:main} is illustrated in Fig.~\ref{fig:main}.
At a generic time interval $K$, the communication of packets to the access point is governed by the choice of the MCW of every user.
In the example, every user has selected the same MCW at time $K$ (depicted by their blue states) but this is not known by node $0$. 
Instead, node $0$ observes the operation of the wireless network $\bbo^K$ and, of course, knows its own $\omega_0^K$.
The concatenation of these observations for the last $M$ time intervals is used as the input to a trained neural network (Rainbow DQN).
The output of this neural network consists of a vector containing the Q-values corresponding to every possible action (MCW) that node $0$ can select.
Node $0$ will select as $\omega_0^{K+1}$ the MCW that attains the largest Q-value at the output of the DQN [cf.~\eqref{E:opt_strategy}], while $\omega_i^{K+1}$ for $i \neq 0$ evolves following an unknown stochastic process.
This same process is repeated and node $0$ relies again on the neural network to select $\omega_0^{K+2}$, and so on.
As we illustrate in Section~\ref{sec:results} for several settings, the discounted sum of utilities yielded by the trained rainbow DQN with optimal parameters $\theta^*$ is comparable with the best attainable strategy when full knowledge of the system is available.

In order to find the optimal parameters $\theta^*$ for our neural network in the first place, we rely on the established strategy of experience replay training; see Section~\ref{sec:results} for implementation details.
In a nutshell, inspired by the fixed-point solution of the classical iteration in~\eqref{E:value_iteration}, we update the values of $\theta$ via gradient descent to minimize the following loss
\begin{equation}
 \mathcal{L}(\theta) = {\mathbb{E}}_{s'} \left( {Q_\theta(s,a)} - \left({r(s,a)+\alpha \max_{a'} Q_\theta(s',a)}\right) \right)^2.
 \label{eqn:loss}
\end{equation}
It should be noted that, in our implementation, instead of using a simple multi-layer perceptron to approximate the optimal Q-function, we rely on recent modifications that have shown empirical improvements in terms of training stability and amount of data needed for training.
Specifically, the rainbow agent~\cite{hessel2017rainbow} incorporates six improvements (double DQN, prioritized replay, duelling networks, multi-step learning, distributional RL, and noisy nets) that led to superior performance over vanilla DQN in gaming environments.
We have also witnessed those benefits in the implementation of deep Q-learning for CW design.

\section{Numerical experiments}
\label{sec:results}

    
To evaluate the performance of the proposed approach, we simulate a wireless network using the NS3 simulator. 
Nodes in the network are modelled to follow the IEEE 802.11b protocol with a constant data rate of 1 Mbps. 
Furthermore, a constant packet size of 1400 bytes is used and the observation interval length T is fixed as 20 seconds.
Unless otherwise stated, the considered networks have $N=10$ nodes.
    
We use two coupled feed forward neural networks as our rainbow DQN, each having four layers with 32 nodes per layer and ReLU non-linearities.\footnote{Code to replicate the numerical experiments here presented can be found at \href{https://github.com/kumarabhish3k/Rainbow-DQN-for-Contention-Window-design.git}{https://github.com/kumarabhish3k/Rainbow-DQN-for-Contention-Window-design.git}.}
For all experiments, the discount factor is set as $\alpha=0.9$ and the DQN is trained for 5000 episodes. 
At the start of each episode, node 0 chooses an action uniformly from the set of actions and the trajectory evolves for a specified number of time steps. 
We fix the number of time steps in each episode to be equal to 50. 
Rainbow DQN is implemented using the PyTorch~\cite{pytorch} library in Python and training is done using mini-batch gradient descent with a batch size of 32. The buffer size for experience replay is set to 10000.
    
\begin{figure*}[!th]
    \centering
    \begin{subfigure}[b]{0.242\textwidth}
        \includegraphics[height=0.72\linewidth]{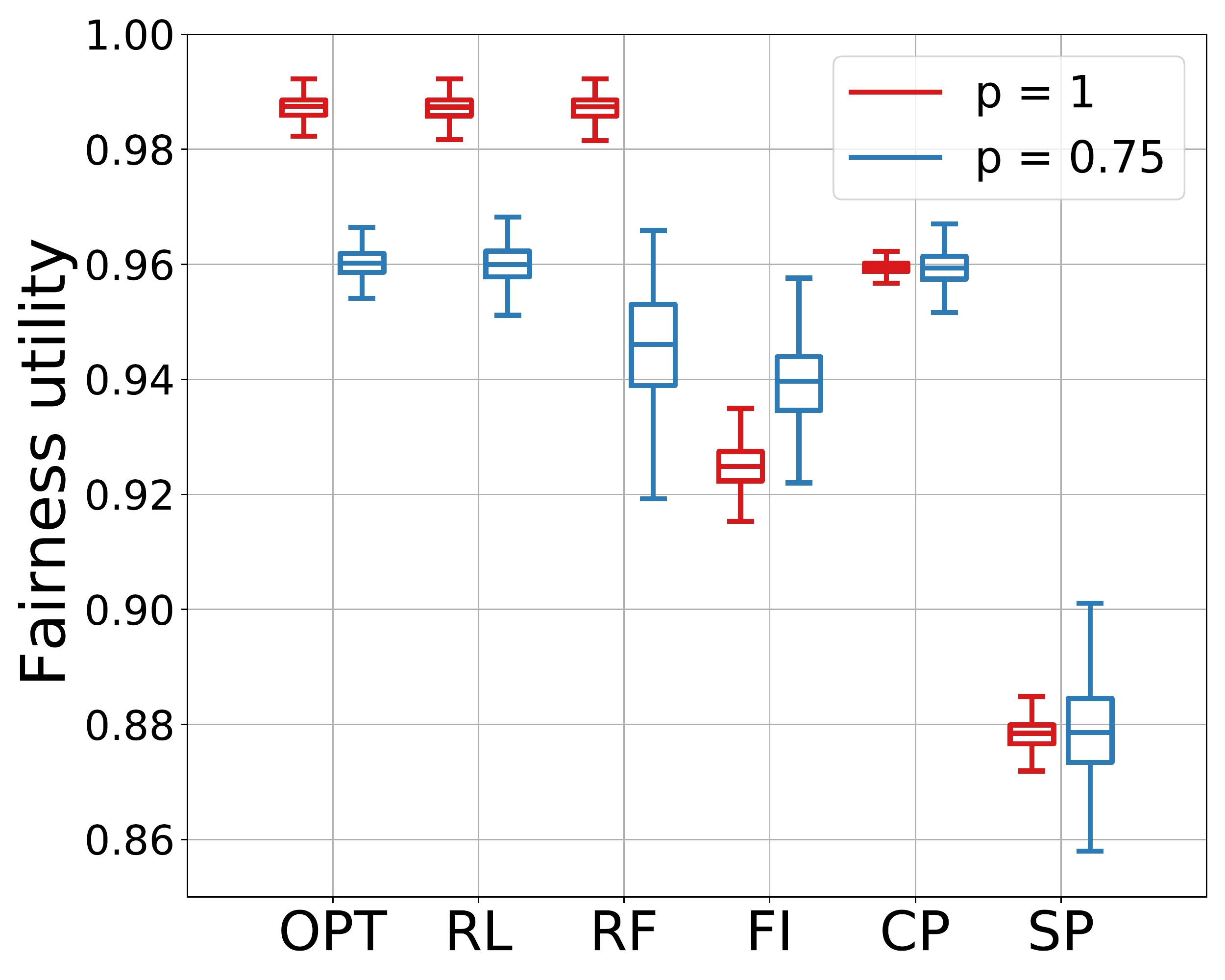}
        \caption{}
        \label{fig:resultMarkov}
    \end{subfigure}%
~     
    \begin{subfigure}[b]{0.242\textwidth}
        \includegraphics[height=0.72\linewidth]{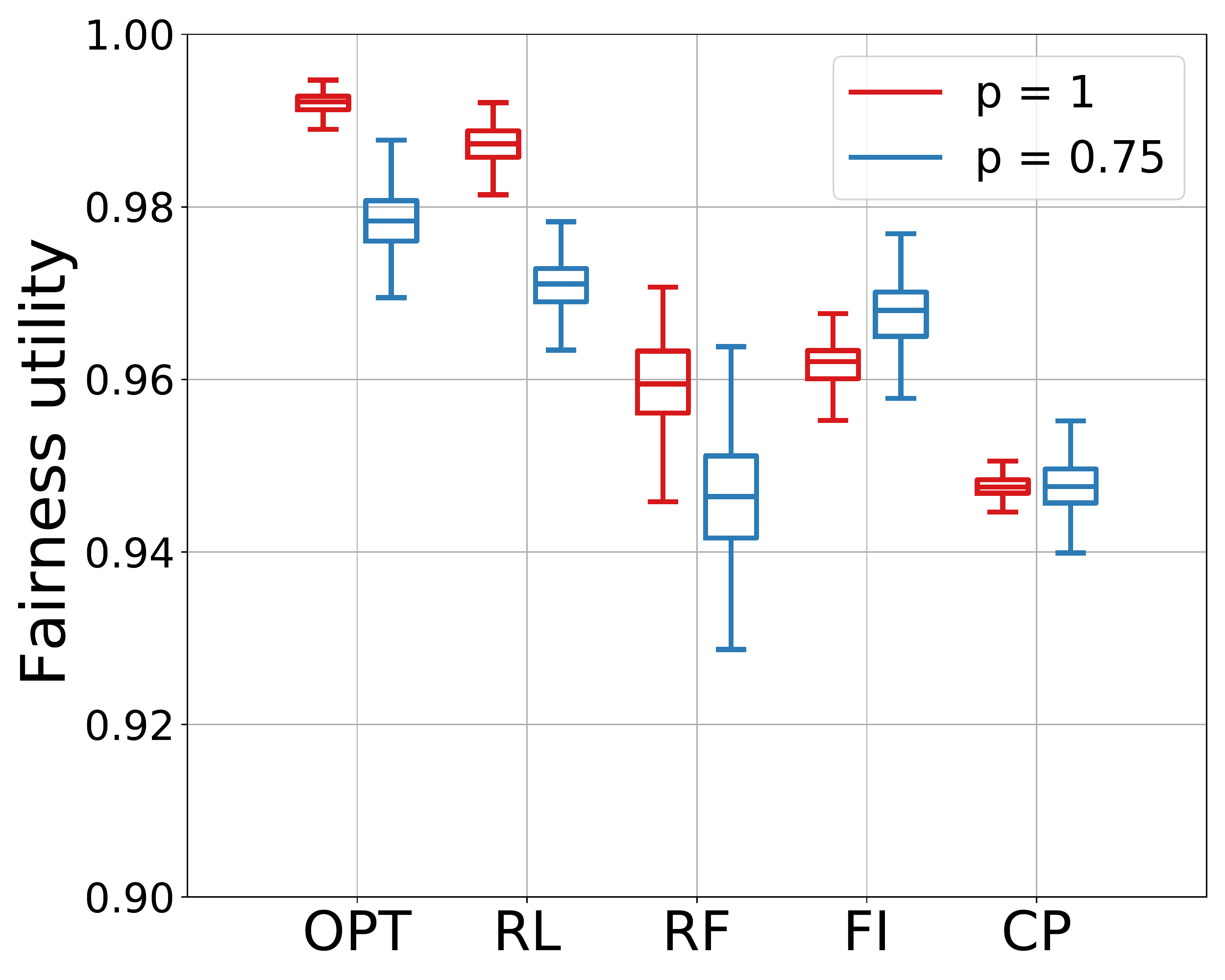}
        \caption{}
        \label{fig:resultNonMarkov1}
    \end{subfigure}  
~
    \begin{subfigure}[b]{0.242\textwidth}
        \includegraphics[height=0.72\linewidth]{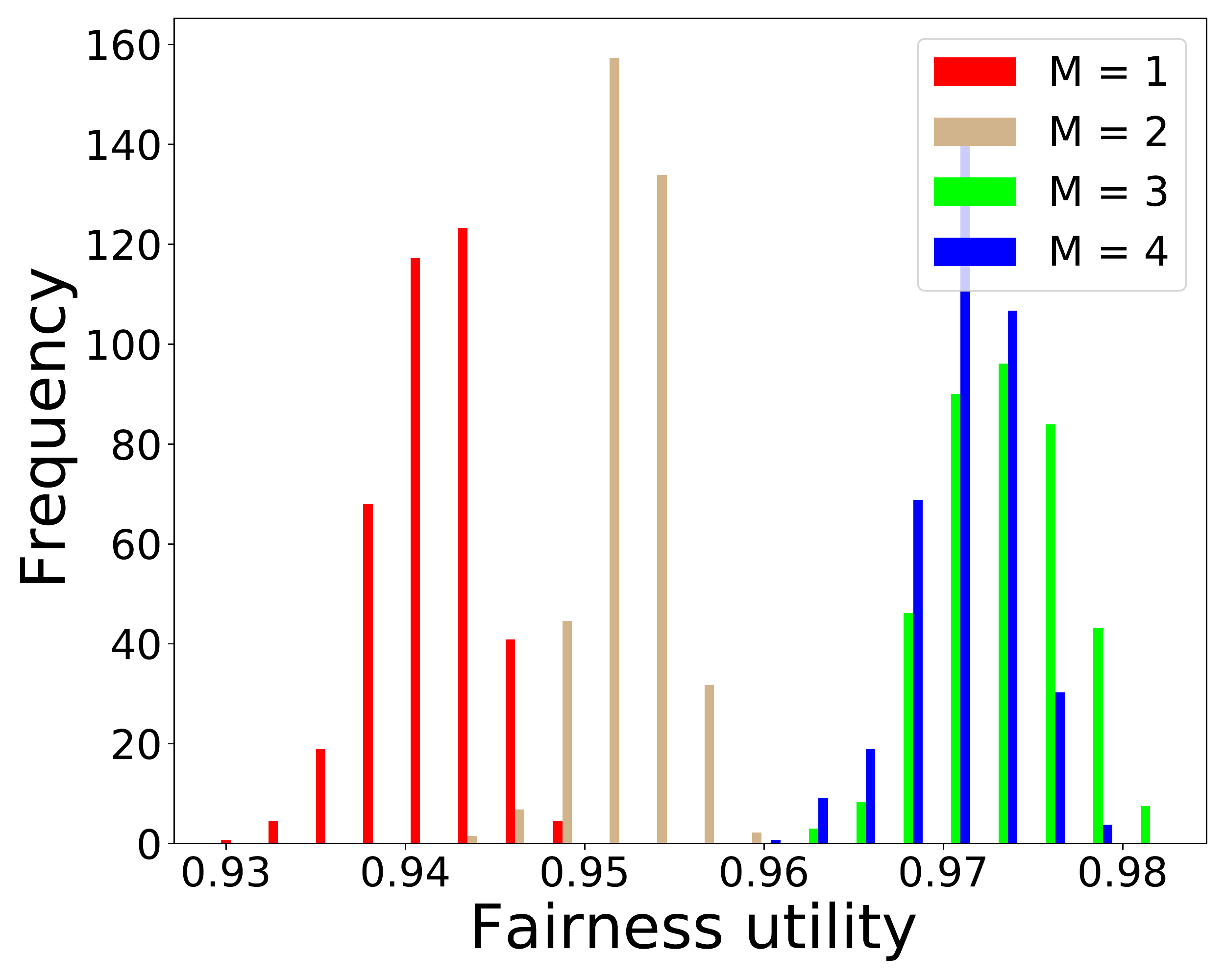}
        \caption{}
        \label{fig:resultNonMarkov2}
    \end{subfigure}%
~
    \begin{subfigure}[b]{0.242\textwidth}
        \includegraphics[height=0.72\linewidth]{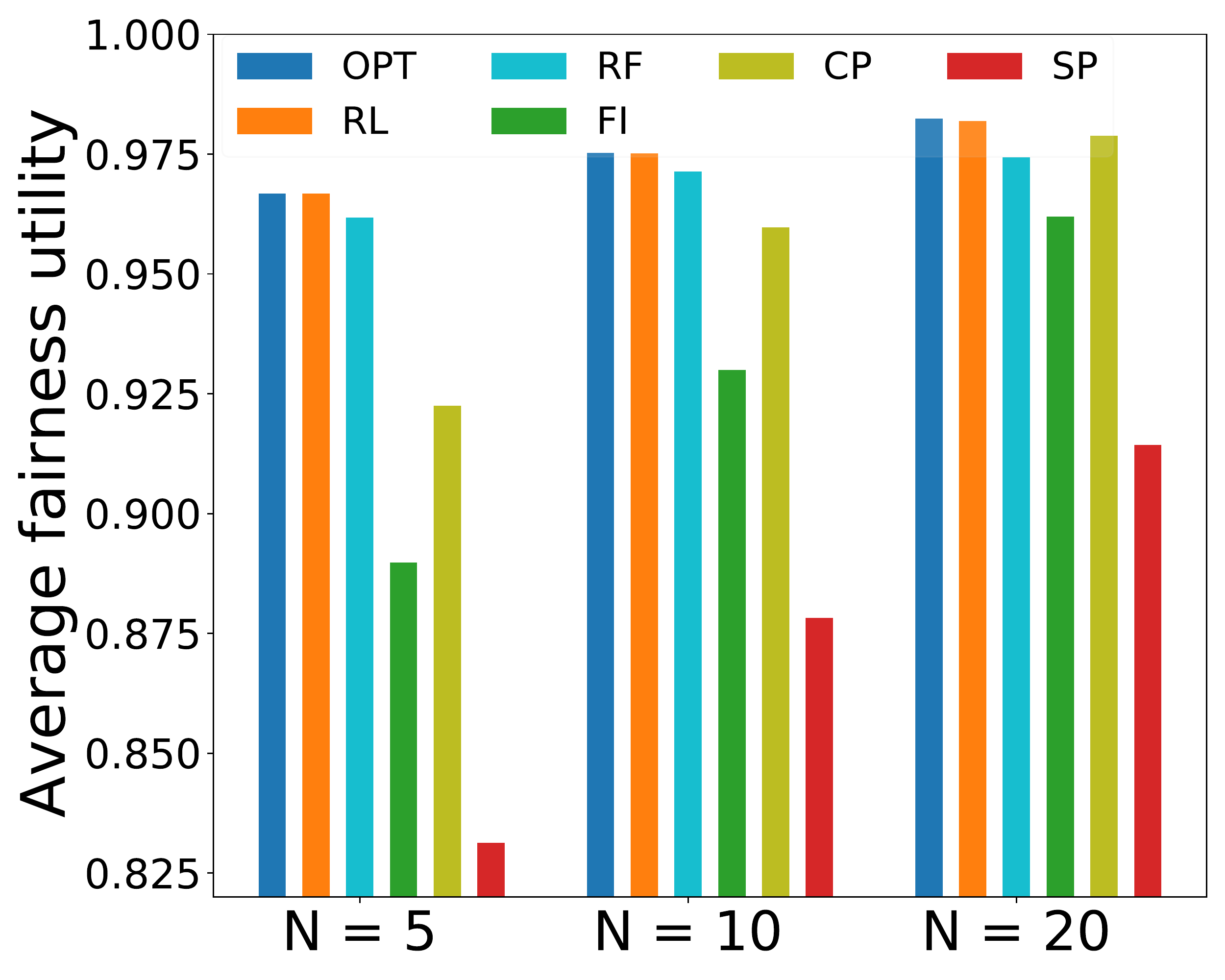}
        \caption{}
        \label{fig:resultsize}
    \end{subfigure}%
    \caption{Performance evaluation of the proposed RL method. 
    (a)~Boxplots for fairness utility over 500 episodes of 50 time steps for the six methods under consideration. A Markov process with $p \in \{0.75,1\}$ is considered for the evolution of the MCW of other nodes. 
    (b)~Counterpart of (a) for the more complex stochastic process described in the text. SP is not shown due to its low relative performance, achieving mean fairness utilities of 0.752 for both values of $p$.
    (c)~Histograms of the utility attained by the RL agent for different lengths of the memory buffer $M$. The respective mean utilities yielded by the increasing values of $M$ are $0.941$, $0.953$, $0.973$, and $0.971$. 
    (d)~Average network utility over 500 episodes for varying number of nodes $N \in \{5,10,20\}$.}
    \label{fig:results}    
\end{figure*}
In evaluating the performance of our proposed RL approach, we compare it with the following five baselines:

\vspace{1mm}
\noindent i) \textbf{Optimal design (OPT):} 
This benchmark assumes full knowledge of the underlying stochastic process and computes explicitly the expected value in Problem~\ref{P:main}. 
Although unrealistic in practice, it sets the upper bound of what is achievable by other methods.

\noindent ii) \textbf{Random forest (RF)~\cite{abyaneh2019intelligent}:} The classifier is trained using the local observations $\{\bbo^k, \omega_0^k\}_{k=K-M+1}^K$ as input and $\omega_0^{*{K+1}}$ as target labels. 
The training labels $\omega_0^{*{K+1}}$ are selected as the optimal actions that maximize the fairness utility at the next time step.
We fix the number of trees as 15 and the depth of each tree as 5.  

\noindent iii) \textbf{Fairness index approach (FI)~\cite{bensaou2000fair}:}
Node $0$ computes a local fairness index and follows a threshold-based policy to update its MCW. 
We set this threshold as $C=1.5$; see~\cite{bensaou2000fair} for details.

\noindent iv) \textbf{Optimal constant policy (CP):} Just like for OPT, we assume complete knowledge of the system. 
However, we restrain the design of the MCW to be constant, i.e., it cannot change across time steps.
Improvements upon CP illustrate the value of an adaptive design.

\noindent v) \textbf{Standard protocol (SP):}
Node 0 follows a static policy as mentioned in the IEEE 802.11 protocol with its MCW fixed at 32.

\vspace{1mm}
Before proceeding with the evaluation, it should be noted that FI is not a predictive method but rather a reactive one in the sense that, whenever the system deviates from fairness, node $0$ would react and try to restore a fair setting.
Thus, it is bound to underperform in a fast-changing setting.
For a fairer comparison, we consider the case where the update speed of FI is three times that of other methods, i.e., FI updates its MCW thrice in every time slot as opposed to once for the other methods (and zero times for CP and SP).

We first consider a simple scenario where the MCWs of all other nodes follow a Markov process with only two states. In state $1$, $\omega_i = 32$ for all $i \neq 0$ and, in state $2$, $\omega_i = 128$ for all $i \neq 0$.
At each discrete time step, the system switches states with probability $p$, and we analyze the cases where $p \in \{0.75, 1\}$.
The action space of node $0$ (potential MCW choices) is $\ccalA = \{32,48,64,96,128\}$.
Since the underlying process is Markovian, we consider a memoryless implementation with $M=1$ (cf. Fig.~\ref{fig:main}).
Fig.~\ref{fig:resultMarkov} reveals that, when $p=1$, RL and RF achieve optimal performance even in the absence of knowledge of the underlying Markov process. 
The advantage of RL over the supervised RF cannot be appreciated in this deterministic process.
The lack of adaptability of CP and SP makes them lag behind and this fast-changing setting is not well suited for FI even with updates three times faster.
This latter effect is attenuated when $p=0.75$ since the system tends to remain longer in a given state.
More importantly, for $p=0.75$, RL is still able to learn the optimal policy while the performance of RF is degraded.
As a consequence of the stochasticity of the process, similar observations $\{ \bbo^k, \omega_0^k\}$ can lead to drastically different optimal values for $\omega_0^{*{k+1}}$, thus confusing the supervised RF classifier.

As our second evaluation setting, we consider the case where the MCWs of others follow a more complex process.
More precisely, we consider five states $\{s_j\}_{j=1}^5$ where $\omega_i = 2^{j-1}  32$ for all $i \neq 0$ at state $j$.
At each time step, the system transitions to the next state with probability $p$, however, the next state follows an increasing trajectory from $s_1$ to $s_5$ followed by a decreasing trajectory from $s_5$ to $s_1$, and so on.
In this sense, the next state does not only depend on the current state but also on the previous one.
To incorporate memory in the predictive methods RL and RF, we consider $M=4$.
Furthermore, we consider the action space $\ccalA = \{ 2^{j-1}  32 \}_{j=1}^5 \cup \{ 2^{j-1}  48 \}_{j=1}^4$.
Fig.~\ref{fig:resultNonMarkov1} reveals that RL achieves the closest utility to optimal among all methods considered for both values of $p$.
As expected, the scenario where $p=0.75$ is more challenging except for the reactive method FI that benefits from slower changes in MCWs. 
Also notice that the performance gap between OPT and RL is reasonable since the former has perfect knowledge of the underlying system whereas the latter must rely exclusively on local observations to gather understanding of the underlying dynamics.
The absence of this gap in Fig.~\ref{fig:resultMarkov} can be attributed to the simplicity of the model and the smaller size of~$\ccalA$. 

Selecting $M>1$ is essential for the RL agent to meaningfully learn from experience in this more complex setting and, for the cases where $p<1$, having $M>2$ can also be beneficial.
This is illustrated in Fig.~\ref{fig:resultNonMarkov2}, where we present the histograms of the fairness utility attained by RL for different values of $M$ over 500 episodes for the case $p=0.75$.
There is a big increase in performance when going from $M=1$ to $M=2$ in accordance with the memory of the underlying system.
Moreover, a similar jump is observed when going from $M=2$ to $M=3$.
This is as expected since, for this value of $p$, the transition to the next state might be delayed underscoring the value of longer memory.
The marginal gain for $M=4$ decreases and, in fact, a small loss is observed which can be attributed to the fact that a larger model must be fit (larger dimension of input to the rainbow DQN) without apparent modeling gain of the environment.

Finally, we consider the Markov process analyzed in our first experiment but for $p = 0.9$ and vary the number of nodes $N \in \{5,10,20\}$.
In Fig.~\ref{fig:resultsize} we portray the utility attained by the six considered methods averaged over 500 episodes as a function of the network size. 
It can be observed that RL achieves utility closest to optimal for all network sizes. 
As one might expect, the effect that an intelligent node has in promoting fairness gets diluted in larger networks, thus, resulting in a smaller dynamic range of utilities across methods when $N=20$.
In particular, the constant conservative strategy of CP that selects a large MCW ($\omega_0^k=128$ for all $k$) attains a small suboptimality gap, but does not match the RL performance. 
Note that RL outperforms the standard protocol for all values of $N$.
Indeed, the consistent near-optimality of the proposed RL approach across network sizes is encouraging and motivates a range of related future work.

\section{Conclusions and future work}
\label{sec:conclusions}
We formulated the design of CWs in wireless networks as an RL problem. 
Within this framework, we leveraged a state-of-the-art DQN architecture to select the optimal value of the MCW in a sequential manner exclusively from local observations. 
Through NS3 simulations, we showed that the proposed RL agent achieves close-to-optimal behavior and outperforms other learning-based and rule-based methods.
Ongoing work includes: i)~the implementation of RL for parameter learning on other access protocols, ii)~the combination of RL with graph neural networks to accommodate the case where every node is intelligent and can be trained based on its local observations, and iii)~the incorporation of more realistic topologies that go beyond star connections to a common access point.

\newpage

\bibliographystyle{newIEEEbib}
\bibliography{main}

\end{document}